\documentclass[global]{svjour}
\usepackage{epsfig}

\def\U{{{\cal U}}}
\def\F{{{\cal F}}}
\def\S{{{\cal S}}}

\title{Enhancing Histograms by Tree-Like Bucket Indices\thanks{An
    abridged version of this paper appeared in the Proceedings
    of the International Conference on Data Engineering (ICDE 2002),
    IEEE Computer Society 2002, ISBN 0-7695-1531-2 \protect\cite{Buccafurri02Improving}}
}

\author{Francesco Buccafurri \and\inst{1}
    Gianluca Lax \and\inst{1}
    Domenico Sacc\`a \and\inst{2}
    Luigi Pontieri \and\inst{2}
    Domenico Rosaci\inst{1}
}

\institute{DIMET Dept., University ``Mediterranea'' of Reggio Calabria, Italy \\
    \email{\{bucca,lax,domenico.rosaci\}@unirc.it}
    \and
    DEIS Dept., University of Calabria, \& ICAR-CNR, Rende, Italy \\
    \email{pontieri@icar.cnr.it, sacca@unical.it}
}

\begin{document}
\maketitle
\begin{abstract}
Histograms are used to summarize the contents of relations into a
number of buckets for the estimation of query result sizes.
Several techniques (e.g., MaxDiff and V-Optimal) have been
proposed in the past for determining bucket boundaries which
provide accurate estimations. However, while search strategies for
optimal bucket boundaries are rather sophisticated, no much
attention has been paid for estimating queries inside buckets and
all of the above techniques adopt naive methods for such an
estimation. This paper focuses on the problem of improving the
estimation inside a bucket once its boundaries have been fixed.
The proposed technique is based on the addition, to each bucket,
of 32-bit additional information (organized into a 4-level tree
index), storing approximate cumulative frequencies at 7 internal
intervals of the bucket. Both theoretical analysis and
experimental results show that, among a number of alternative ways
to organize the additional information, the 4-level tree index
provides the best frequency estimation inside a bucket. The index
is later added to two well-known histograms, MaxDiff and
V-Optimal, obtaining the non-obvious result that despite the
spatial cost of 4LT which reduces the number of allowed buckets
once the storage space has been fixed, the original methods are
strongly improved in terms of accuracy.
\end{abstract}


\section{Introduction}
A {\em histogram} is a lossy compression technique used for
representing efficiently a relation. It is based on the partition
of one of the relation attributes into {\em buckets} and the
storage, for each of them, of a few summary information in place
of the detailed one. Among others, some important examples of
application domains of histograms are the estimation of query
selectivity
\protect\cite{IoPo95,Jagadish98Optimal,Poosala96Improved,Ja*01,Wu03Using},
temporal databases, where histograms are used for improving the
join processing \protect\cite{Sitzmann00Improving}, statistical
databases, where histograms represent a method for approximating
probability distributions \protect\cite{Malvestuto93Universal}.
Recently, histograms have received a new deal of interest, mainly
because they can be effectively used for approximating query
answering in order to reduce the query response time in on-line
decision support systems and OLAP \protect\cite{Poosala99Approx},
as well as the problem of reconstructing original data from
aggregate information \protect\cite{BuFuSa01} and, finally, in the
context of Data Streams
\protect\cite{Guha01Data,Babcock02Models,Datar02Maintaining,Guha02Histogramming}.

For a given storage space reduction, the problem of determining
the best histogram is crucial. Indeed, different partitions lead
to dramatically different errors in reconstructing the original
data distribution, especially for skewed data. To better explain
the problem, consider a typical case of recovering original data
from a histogram: the evaluation of range queries. Think to a
histogram defined on the attribute $X$ of a relation $R$ as a set
of non-overlapping intervals of $X$ covering all values assumed by
$X$ in $R$. To each of these intervals, say $B$, the number of
occurrences (called {\em frequency}) in $R$, having the value of
$X$ belonging to the interval $B$, is associated (and included
into a data structure called {\em bucket}). A {\em range query},
defined on an interval $Q$ of $X$, evaluates the number of
occurrences in $R$ with value of $X$ in $Q$. Thus, buckets embed a
set of pre-computed disjoint range queries capable of covering the
whole active domain of $X$ in $R$ (with active here we mean
attribute values actually appearing in $R$). As a consequence, the
histogram does not give, in general, the possibility of evaluating
exactly a range query not corresponding to one of the pre-computed
embedded queries. In other words, while the contribution to the
answer coming from the sub-ranges coinciding with entire buckets
can be returned exactly, the contribution coming from the
sub-ranges which partially overlap buckets can be only estimated,
since the actual data distribution inside the buckets is not
available.

It turns out that it is convenient to define the boundaries of
buckets in such a way that the estimation error of the
non-precomputed range queries is minimized (e.g., by avoiding that
large frequency differences arise inside a bucket). In other
words, among all possible sets of pre-computed range queries, we
find the set which guarantees the best estimation of the other
(non-precomputed) queries, once a technique for estimating such
queries is defined. This issue is being investigated since some
decades, and a large number of techniques for arranging histograms
have been proposed
\protect\cite{Cri81,Cri84,IoPo95,Jagadish98Optimal,Poosala96Improved,DonIoa00,Ja*01}.

All these techniques adopt simple methods for estimating
non-precomputed queries (actually, their portions partially
overlapping buckets). The most significant approaches are the {\em
continuous value assumption} (often denoted in this paper by CVA)
\protect\cite{Sac79}, where the estimation is made by linear
interpolation on the whole domain of the bucket, and the {\em
uniform spread assumption} (denoted by USA)
\protect\cite{Poosala96Improved}, which assumes that values are
located at equal distance from each other so that the overall
frequency sum can be equally distributed among them.

An interesting problem is understanding whether, by exploiting
information typically contained in histogram buckets, and possibly
adding a few summary information, the frequency estimation inside
buckets, and then, the histogram accuracy, can be improved. This
paper focuses on this problem. Starting from the consideration of
limits of CVA and USA studied in \protect\cite{BuFuSa01}, we
propose to use some additional storage space in order to describe
the distribution inside a bucket in an approximate yet very
effective way.

The first step is studying how to use these 32 additional bits in
order to maximize benefits in terms of accuracy. Our analysis
shows that the trivial technique of partitioning the bucket into 8
equal-size parts and encoding each corresponding sum by 4 bits,
leads to high scaling errors since it is needed to represent each
sum as a fraction of the overall sum of the bucket. Our proposal
then relies on the idea of storing partial sums internal to the
bucket in a hierarchical fashion, using a tree-like index
(occupying 32 bits). This way, the sum contained in a given tree
node, can be represented as a fraction of the sum contained in the
parent node, which is a value (reasonably) smaller than the
overall sum of the bucket. It turns out that the encoding length
may decrease as the level of the tree increases. The benefits we
expect by applying this approach concern the scaling error. But a
crucial point is to decide how to arrange the tree, that is, how
far going down in depth with the index. Of course, the higher the
resolution, the larger the number of embedded precomputed range
queries (internal to the buckets) is. Hence, we expect better
accuracy as the resolution increases. However, increasing
resolution reduces the number of bits available for encoding
nodes, and, thus, amplifies scaling errors. We study the above
trade-off by considering the two possible (from a practical point
of view) tree-indices with 32 bits, which we call 3LT and 4LT,
with depth 3 and 4, respectively. The analysis leads to the
conclusion that the 4LT-index represents the best solution.

The next step is then understanding whether this improvement of
accuracy for the estimation inside buckets can really give
benefits in terms of accuracy of a histogram arranged by one of
the existing techniques. This problem is not straightforward:
think, to mention the most evident aspect, that 4LT buckets use 32
bits more than CVA ones, and, then, for a fixed storage space,
allows a smaller number of buckets. The last part of this paper is
thus devoted to evaluate the effects of the combination of the 4LT
technique with existing methods for building histograms. Through a
deep experimental comparative analysis conducted, for a fixed
storage space, over several data sets, both synthetic and
real-life, we show that 4LT improves significantly the accuracy of
the considered histograms. Therefore this paper, beside giving the
specific contribution of proposing a technique (i.e., the 4LT) for
estimating accurately range queries internal to buckets, proves
the more general result that going beyond classical techniques
(i.e., CVA and USA) for the estimation inside buckets may give
concrete improvements of histogram accuracy.

It is worth noting that the choice of MaxDiff and V-Optimal
histograms for testing our method does not limit the generality of
the 4LT index, which is applicable to every bucket-based
histogram\footnote{There are histograms, like wavelet-based ones,
that are not based on a set of buckets.}. Nevertheless, it is not
limited the validity of our comparison, since MaxDiff and
V-Optimal, despite their non-young age, are still considered in
this scientific community as point of references due to their
accuracy \protect\cite{Ioannidis03History}.

The paper is organized as follows. In Section
\ref{sec-preliminary}, we introduce some preliminary definitions.
The comparison, both experimental and theoretical, among a number
of techniques including our tree-based methods (3LT and 4LT) for
estimating range queries {\em inside} a bucket is reported in
Section \ref{sec-Estimation}. Therein, 3LT and 4LT are also
presented. From this analysis it results that 4LT has the best
performances in terms of accuracy. Thus, 4LT can be combined to
every bucked-based histogram for increasing its accuracy. Section
\ref{sec-Improved} presents a large set of experiments, conducted
by applying 4LT to two, well-known methods, {\em MaxDiff} and {\em
V-Optimal} \protect\cite{Poosala96Improved}. Results show high
improvements in the estimation of range queries w.r.t. to the
original methods --- of course, the comparisons are made at parity
of storage consumption so that the revised methods use less
buckets to compensate the additional storage for the 4LT indices.
The 4LT technique provides good results also when combined with
the very simple method {\em EquiSplit}, which consists in dividing
the histogram value domain into buckets of the same size so that
the bucket boundaries need not to be stored, thus obtaining a very
high number of buckets at the same compression rate. We draw our
conclusions in Section \ref{sec-Conclusion}.

\section{Basic Definitions}\label{sec-preliminary}

Given a relation $R$ and  an attribute $X$ of $R$, a histogram for $R$ on
$X$ is constructed as follows. Let  $\U = \{u_1, ... , u_m\}$ be
the set of all possible values (the {\em domain}) of $X$ and let
$u_i < u_{i+1}$, for each $i$, $1 \leq i <m$. The {\em frequency
set} for $X$ is the set $\F =\{f(u_1), ... , f(u_m) \} $ such that
for each $i$, $1 \leq i \leq m$, $f(u_i)$ is the number of
occurrences of the attribute value $u_i$ in the relation $R$. The
{\em cumulative frequency set} $\S =\{s_1, ... , s_m \}$ contains
the value $s_i = \sum_{j=1}^i f(u_j)$ for each attribute value
$u_i$. The {\em value set} $V=$ $\{u_i \in \U \ | \ f(u_i) > 0 \}$
is the active domain of $X$ in $R$ as it consists of all attribute
values actually occurring in the relation $R$ ({\em non-null
values}). Given any $u_i$ in $V$, the {\em spread} $d_i$ of $u_i
\in V$ for $1 \leq i < n$  is defined as 1 if $u_i$ is the last
non-null value or otherwise as the difference $u_j - u_{i}$, where
$u_{j}$ is the first non-null value for which $u_j > u_i$ (i.e.,
$d_i$ is the distance from $u_i$ to the next non-null value).

A {\em bucket} $B$ for $R$ on $X$  is a 4-tuple $\langle inf, sup,
t, c \rangle$, where $u_{inf}$ and $u_{sup}$, $1 \leq inf \leq sup
\leq m$, are the boundaries of the domain range pertaining to the
bucket, $t$ is the number of non-null values occurring in the
range, and $c = \sum_{i=inf}^{sup} f(u_i)$ is the sum of
frequencies of all values in the range.
We say that the bucket $B$ is {\em
1-biased} if $u_{sup}$ is not null; if also $u_{inf} $ is not
null, then we say that $B$ is {\em 2-biased}.

A {\em histogram} $H$ for $R$ on $X$  is a $h$-tuple $\langle
B_1,B_2, ..., B_h \rangle$ of buckets such that: (1) for each $1
\leq i < h$, the upper bound of $B_i$ precedes the lower bound of
$B_{i+1}$ and (2) $u \in V$ implies $u \in B_i$, for some $i$, $1
\leq i \leq h$. Condition (1) guarantees that buckets do not
overlap each other, and condition (2) enforces that every non-null
value be hosted by some bucket. Classically, histograms have
2-biased buckets; sometime, for storage optimizations, 2-biased
buckets are made 1-biased by replacing the lower bound of each
bucket with the successive in the domain of the upper bound of the
preceding bucket.

A classical problem on histograms is: given a histogram $H$ and a
(range) query of the form $u_j \leq X \leq u_i$, $1 \leq j \leq i \leq m$,
estimate the overall frequency $\sum_{k=j}^i f(i)$ in the range from $u_j$
to $u_i$.

\section{Estimation Inside a Bucket}\label{sec-Estimation}

In this section we deeply investigate the problem of frequency
estimation inside buckets. First of all,  we present the classical
two techniques (CVA and USA), discuss their limitations and
propose some simple alternatives. Then we introduce a novel
technique which is based on a 4-level tree index storing
approximate representations of the partial sums of 7 fixed bucket
intervals. Later we evaluate the accuracy of the various
techniques by performing both a theoretical analysis of errors and
a number of experiments on some typical sample distributions.

\subsection{Notations and Problem Formulation}

Let $B=\langle inf, sup, t, c \rangle$ be a bucket on an attribute
$X$ of a relation $R$. Without loss of generality, we assume that
$inf=1$ and $sup=b$ so  that we can represent the frequency set
inside the bucket as a vector $F$ with indexes ranging from $1$ to $b$
({\em frequency vector of} $B$). Similarly, the cumulative
frequencies are represented by a vector $S$ with indexes from $1$
to $b$ ({\em cumulative frequency vector of} $B$). Hence, for each
$i$, $1 \leq i \leq b$, $F[i]\geq 0$ is the frequency of the value
$u_i$ while $S[i]=$ $\sum_{j=1}^{i}F[j]$ is the cumulative
frequency. Then $c=S[b]$ is the sum of all frequencies in the
bucket; moreover, for notation convenience, we assume that
$S[0]=0$.

The problem of the estimation inside a bucket can be formulated as
follows: {\em given any pair} $i,j$, $1 \leq i \leq j \leq b$, such
that $d=j-i+1 < b$, {\em estimate the range query} $S[j] - S[i-1] =$
$\sum_{k=i}^j
F[k]$.
We focus our attention on the basic problem of estimating $S[d]$
(then by assuming $i=1$).

We introduce now the following notation.
Given $1 \leq i \leq j \leq 8$,
we denote by $\delta_{i/j}$ the sum
$\sum_{i=x}^{y} F[i]$,
where $x=1+ \lceil \frac{b}{j}  \cdot (i -1) \rceil$
and $y= \lceil \frac{b}{j}  \cdot i\rceil$.
$\delta_{i/j}$ represents the frequency sum of the $i-$th
elements of the partition of $B$ into $j$ equal size sub-ranges.
Thus, the frequency sum for a bucket is $\delta_{1/1}$; the
frequency sums for two halves are $\delta_{1/2}$ and
$\delta_{2/2}$; the frequency sums for the 4 quarters are
$\delta_{i/4}$, $1 \leq i \leq 4$; the frequency sums for the 8
eighths are $\delta_{i/8}$, $1 \leq i \leq 8$, and so on.

\subsection{Estimation Techniques}

Next we illustrate the existing approximation techniques
and discuss some additional simple approaches.

\noindent{\bf Continuous Value Assumption (CVA).} The estimation
of $S[d]$ is computed as $\widetilde{S}[d]=\frac{d}{b} \cdot c$.
In words, the partial contribution of a bucket to a range query
result is estimated by linear interpolation. As pointed out in
\protect\cite{Buccafurri99Compressed,BuFuSa01}, the above
estimation coincides with the expected value of the $S[d]$ when it
is considered a random variable over the population of all
frequency distributions in the bucket for which the overall
cumulative frequency is $c$. \noindent{\bf Uniform Spread
Assumption (USA).} The estimation of $S[d]$ is given by
$\widetilde{S}[d] = \left ( 1 + \frac{(t-1)\cdot (d-1)}{(b-1)}
\right ) \cdot \frac{c}{t}$, where $t$ is the number of non-null
attribute values in the bucket. The uniform spread assumption
assumes that such values are distributed at equal distance from
each other and the overall frequency sum is equally distributed
among them. Obviously, in this case the information $t$ is
necessary.  We stress that, as discussed in
\protect\cite{BuFuSa01}, this estimation is not supported by any
unbiased probabilistic model so the assumption is rather
arbitrary.

\noindent{\bf 1-Biased Estimation (1b).} The possibly available
information on the number $t$ of non-null elements cannot be
exploited in the estimation unless some further information on the
frequency distribution is either available or assumed (as for the
USA estimation). We next show how to exploit the fact that a
bucket is often 1-biased (i.e., $u_b$ is not null) using the
probabilistic approach proposed in \protect\cite{BuFuSa01}. This
approach assumes that the query is a random variable on the
population  of all 1-biased frequency distributions having $c$ as
overall cumulative frequency. The estimation of the range query
$S[d]$ for a 1-biased bucket is given by $\widetilde{S}[d]=
\frac{d}{b-1} \cdot \frac{t-1}{t} \cdot c$.

\noindent{\bf 2-Split Estimation (2s).} We split the bucket into
two parts of the same size and  store the cumulative frequency of
the first part, say $\delta_{1/2}=S[b/2]$ ---  we therefore need
additional storage space (typically 32 bits). We call this method
{\em 2-split} or $2s$ for short. Following this approach, the
estimation of the range query $S[d]$ is given by
$2 \cdot \frac{d}{b} \cdot  \delta_{1/2}$ if $d \leq \frac{b}{2}$,
$\delta_{1/2} + 2 \cdot \frac{d - b}{b} \cdot (c - \delta_{1/2})$,
otherwise.
Thus we use the CVA techniques for each of the two halves of the
bucket.

\noindent{\bf 4-Split Estimation (4s).} We split the bucket into
4 parts of the same size ({\em quarts}) and  store the
approximate values of the cumulative frequency of the each
part $\delta_{i/4}$, $1 \leq i \leq 4$.
In case the additional available space is 32 bits, we use 8 bits for each
approximate value, which is therefore computed as
$\tilde{\delta}_{i/4}=\langle\frac{\delta_{i/4}}{c}
\times (2^8-1)\rangle$,
where $\langle x \rangle$ stands for $round(x)$.
The frequency sum
for an interval $d$ is estimated by adding the approximate values
of all first quarts that are fully contained in the interval plus
the CVA estimation of the portion of the last eighth that
partially overlaps the interval. Obviously, in order to reduce the
approximation error, in case $d>b/2$, it is convenient to derive
the approximate value from the estimation of the cumulative
frequency in the complementary interval from $d+1$ to $b$.

\noindent{\bf 8-Split Estimation (8s).}
It is analogous to the 4-Split Estimation. The only difference is that the
bucket is
divided into 8 parts ({\em eighths}) and, for each of them, we use
4 bits for storing the cumulative frequency.
Thus, the approximate value of the $i$-th eight ($1 \leq i \leq 4$) , is
computed as
$\tilde{\delta}_{i/8}=\langle\frac{\delta_{i/8}}{c}
\times (2^4-1)\rangle$,
where $\langle x \rangle$ stands for $round(x)$.

\subsection{The Tree Indices for Bucket Frequency Estimation}

We now propose to use 32 bits as sophisticated tree-indices for
providing an {\em approximate description} of the cumulative
frequencies in the bucket --- this index can be easily extended
also to the case that more bits are available. To this end, we store
the approximate value of the cumulative frequency in a suitable number of
intervals
inside the bucket.
The first type of tree-index is 3LT.

\noindent
{\bf 3 Level Tree index (3LT)}
The 3LT index uses 11 bits for
approximating the value of $\delta_{1/2}$, and 10 bits both for
approximating $\delta_{1/4}$ and for $\delta_{3/4}$.

Let $L_{1/2}$ be the
11-bits string corresponding to $\delta_{1/2}$, and let $L_{1/4}$ and
$L_{3/4}$ be the 10-bits strings corresponding, respectively, to
$\delta_{1/4}$ and $\delta_{3/4}$.

The three $L$ strings are constructed as follows:

\vspace{2mm}
\begin{center}
{\small
$L_{1/2} =  \langle\frac{\delta_{1/2}}{\delta_{1/1}}\cdot (2^{11}-1)
\rangle;
\ \ \ \
L_{1/4}= \langle\frac{\delta_{1/4}}{\delta_{1/2}}\cdot (2^{10}-1)\rangle;
\ \ \ \
L_{3/4}= \langle\frac{\delta_{3/4}}{\delta_{2/2}}\cdot (2^{10}-1)\rangle$
}
\end{center}

\vspace{2mm}
\noindent where, we recall, $\langle x \rangle$ stands for $round(x)$.

The approximate values for the partial sums are given by:

\vspace{2mm}
\begin{center}
{\small
$\widetilde{\delta}_{1/1}=\delta_{1/1}=s$\\

$\widetilde{\delta}_{1/2}= \frac{L_{1/2}}{2^{11}-1} \cdot
\widetilde{\delta}_{1/1};
\ \ \ \ \ \ \ \ \ \ \ \ \ \ \ \ \ \ \ \
\widetilde{\delta}_{2/2}= \widetilde{\delta}_{1/1} -
\widetilde{\delta}_{1/2}$ \\

$\widetilde{\delta}_{1/4}= \frac{L_{1/4}}{2^{10}-1}\cdot
\widetilde{\delta}_{1/2};
\ \ \ \ \
\widetilde{\delta}_{2/4}=\widetilde{\delta}_{1/2} -
\widetilde{\delta}_{1/4};
\ \ \ \ \ \
\widetilde{\delta}_{3/4}= \frac{L_{3/4}}{2^{10}-1}\cdot
\widetilde{\delta}_{2/2};
\ \ \ \ \
\widetilde{\delta}_{4/4}=\widetilde{\delta}_{2/2} -
\widetilde{\delta}_{3/4}$\\
}
\end{center}
\vspace{2mm}

Observe that the 32 bits index refers to a 3-level tree whose
nodes store directly or indirectly the approximate values of the
cumulative frequencies for fixed intervals: the root stores the
overall cumulative frequency $c$, the two nodes of the second
level store the cumulative frequencies for the two halves of the
bucket and so on.

\begin{figure}[t]
\epsfig{file=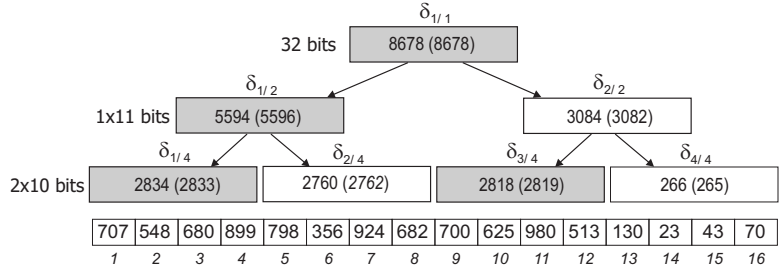,width=11cm}
\caption{The 3-level tree.}\label{fig-3LT}
\end{figure}

\begin{example}\label{3LT-example}
Consider the 3-level tree in Figure
\ref{fig-3LT}. The 32 bits store the following approximate
cumulative frequencies: $L_{1/2}=\langle \frac{5594}{8678} \cdot
2047 \rangle=1320$, $L_{1/4}=\langle \frac{2834}{5594} \cdot 1023
\rangle=518$, $L_{3/4}=\langle \frac{2818}{8678-5594} \cdot 1023
\rangle=935$.
\end{example}

We are now ready to solve the frequency estimation inside the bucket
$B$. Given $d$, $1 \leq d < b$, let $i$ be the integer for which
$\lceil(i-1)/4 \cdot b \rceil \leq d < \lceil i/4 \cdot b \rceil$.
Then the approximate value of $F[d]$ is:

\[
\begin{array}{l}
\widetilde{F}[d]= P(i)+P'(i)+\frac{d-\lceil(i-1)/4 \cdot
b\rceil} {\lceil i/4 \cdot b\rceil-\lceil(i-1)/4 \cdot b\rceil}
\cdot \widetilde{\delta}_{i/4}
\end{array}
\]
\noindent where
\[
\begin{array}{ll}
P(i)= \left \{
\begin{array}{ll}
\widetilde{\delta}_{1/2} & \mbox{if $i > 2$}\\
0 & \mbox{if $i \leq 2$}
\end{array}  \right.
& \ \ \ \ \
P'(i)= \left \{
\begin{array}{ll}
\widetilde{\delta}_{1/4} & \mbox{if $i = 2$}\\
\widetilde{\delta}_{3/4} & \mbox{if $i = 4$}\\
0 & \mbox{otherwise}
\end{array}  \right.
\\
\end{array}
\]
Thus we use the interpolation based on the CVA only inside a
segment of length $\lceil(1/4) \cdot b\rceil$. This component
becomes zero at each distance $d=\lceil i \cdot \frac{b}{4} \rceil$, $1
\leq i < 4$.

32 bits may be distributed in such a way that the granularity of
the tree-index increases w.r.t. 3LT. 4LT index has 4 levels
and uses 6 bits for the first level, 5 bits for the second one and
4 bits for the last level.

\noindent
{\bf 4 Level Tree index (4LT)}
We reserve 4 bits to store the approximate value of each of the
following 4 partial sums: $\delta_{1/8}$, $\delta_{3/8}$,
$\delta_{5/8}$ and $\delta_{7/8}$ --- let $L_{i/8}$, $i=1,3,5,7$,
denote such 4-bits strings. We then use the remaining 16 bits as
follows: the partial sums $\delta_{1/4}$ and $\delta_{3/4}$ are
approximated by the 5-bit strings $L_{1/4}$ and $L_{3/4}$,
respectively, while the partial sum $\delta_{1/2}$ with a 6-bits
string $L_{1/2}$. As a result, the larger the intervals, the higher
is the number of bits used.
The 8 $L$ strings are constructed as follows:

\[
\small
\begin{array}{ll}
L_{1/2} =  \langle\frac{\delta_{1/2}
}{\delta_{1/1}}\cdot (2^6-1)\rangle  &
\\

L_{i/4}= \langle\frac{\delta_{i/4}}{\delta_{j/2}}\cdot (2^5-1)\rangle
& (i=1 \wedge j =1 ), (i=3 \wedge j =2 )\\

L_{i/8}=\langle\frac{\delta_{i/8}}{\delta_{j/4}}\cdot (2^4-1)\rangle
& (i=1 \wedge j =1), (i=3 \wedge j =2), \\

& (i=5 \wedge j =3), (i=7 \wedge j =4)\\

\end{array}
\]

\noindent where,
we recall, $\langle x \rangle$ stands for $round(x)$.

The approximate values for the partial sums are eventually
computed as:

\[
\begin{array}{ll}
\widetilde{\delta}_{1/1}=\delta_{1/1}=c & \\

\widetilde{\delta}_{1/2}= \frac{L_{1/2}}{2^6-1}\times
\widetilde{\delta}_{1/1}\\

\widetilde{\delta}_{2/2}= \widetilde{\delta}_{1/1} -
\widetilde{\delta}_{1/2} &\\

\widetilde{\delta}_{i/4}= \frac{L_{i/4}}{2^5-1}\times
\widetilde{\delta}_{j/2} & (i=1 \wedge j =1), (i=3 \wedge j =2) \\

\widetilde{\delta}_{i/4}=\widetilde{\delta}_{j/2} -
\widetilde{\delta}_{i-1/4} &
(i=2 \wedge j =1),
(i=4 \wedge j =2)\\

\widetilde{\delta}_{i/8}= \frac{L_{i/8}}{2^4-1}\times
\widetilde{\delta}_{j/4} & (i=1 \wedge j =1), (i=3 \wedge j =2) \\ &
(i=5 \wedge j =3), (i=7 \wedge j =4) \\
\widetilde{\delta}_{i/8}=\widetilde{\delta}_{j/4} -
\widetilde{\delta}_{i-1/8} &
(i=2 \wedge j =1),
(i=4 \wedge j =2)\\ & (i=6 \wedge j =3), (i=8 \wedge j =4)
\end{array}
\]

Similarly to the 3LT-index, the 4LT-index
refers to a 4-level tree whose
nodes store directly or indirectly the approximate values of the
cumulative frequencies for fixed hierarchical intervals
starting from the root which stores the
overall cumulative frequency $c$.

\begin{figure}[t]
\epsfig{file=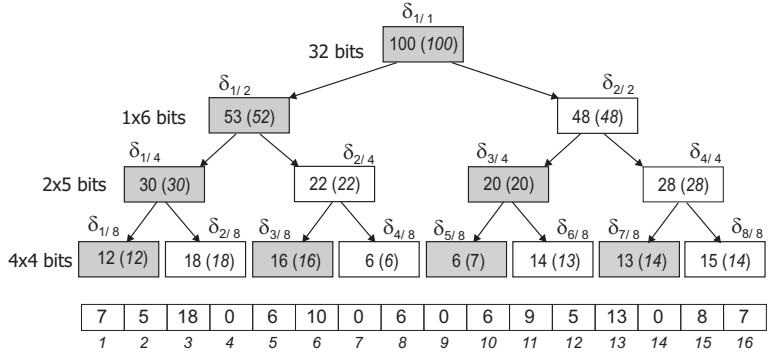,width=11cm}
\caption{The 4-level tree.}\label{fig-4LT}
\end{figure}

\begin{example}
Consider the 4-level tree in Figure
\ref{fig-4LT}.
The 32 bits store the following approximate
cumulative frequencies: $L_{1/2}=33$, $L_{1/4}=18$, $L_{3/4}=13$,
$L_{1/8}=6$, $L_{3/8}=11$, $L_{5/8}=5$, $L_{7/8}=7$.
\end{example}

Again, similarly to the 3LT-index, the frequency estimation inside the
bucket
$B$ can be obtained by exploiting the content of the nodes of the index.
Given $d$, $1 \leq d < b$, and the integer $i$ which
$\lceil(i-1)/8\times b\rceil \leq d < \lceil i/8\times b\rceil$,
the approximate value of $F[d]$ is:
\[
\begin{array}{l}
\widetilde{F}[d]= P(i)+P'(i)+P''(i)+\frac{d-\lceil(i-1)/8\times
b\rceil} {\lceil i/8\times b\rceil-\lceil(i-1)/8\times b\rceil}
\times \widetilde{\delta}_{i/8}
\end{array}
\]
\noindent where
\[
\begin{array}{ll}
P(i)= \left \{
\begin{array}{ll}
\widetilde{\delta}_{1/2} & \mbox{if $i > 4$}\\
0 & \mbox{if $i \leq 4$}
\end{array}  \right.
&
P'(i)= \left \{
\begin{array}{ll}
\widetilde{\delta}_{1/4} & \mbox{if $i = 3,4$}\\
\widetilde{\delta}_{3/4} & \mbox{if $i = 7,8$}\\
0 & \mbox{otherwise}
\end{array}  \right.
\\
\end{array}
\]
\[
P''(i)= \left \{
\begin{array}{ll}
\widetilde{\delta}_{i-1/8} & \mbox{if $i$ is even}\\
0 & \mbox{otherwise}
\end{array}  \right.
\]

Thus we
use the interpolation like in CVA only
inside a segment of length $\lceil(1/8)b\rceil$. This component
becomes zero at each distance $d=\lceil i \times b/8 \rceil$, $1
\leq i < 8$. We call the estimation {\em 4-level tree} or 4LT for short.

\subsection{Worst-case Error Analysis}\label{sec-Analysis}

The approximation error for CVA, 1b, USA and 2s arises only from interpolation.
On the contrary, for other methods (i.e., 4s, 8s, 3LT and 4LT), the scaling
error due to bit saving is added to the interpolation error.
However, all methods but CVA, 1b and USA implement a equi-size division of
the bucket and
3LT and 4LT provide also an index over sub-buckets.
We expect that such a division into sub-buckets produces an improvement
from the side of the interpolation error.
Indeed, sub-buckets increase
the granularity of summarization.
In addition, we expect that index-based methods (i.e., 3LT and 4LT), reduce
the scaling error, since
hierarchical tree-like organization allows us to
represent the sum inside a given sub-bucket, corresponding to a
node of the tree, as a fraction of the sum contained in the parent
node, instead of a fraction of the entire bucket sum (as it happens for the
"flat" methods 4s and 8s).
The worst-case analysis confirms the above observations.
In particular we show that while CVA, 1b and USA are the same, under the
worst-case point of view, 4LT outperforms the other methods.

Results of our analysis are summarized in the following theorem.
Recall that, throughout the whole section, a bucket $B$ of
size $b$ is given.

\begin{theorem}
Let $F$ be the maximum frequency value occurring in $B$ and
let assume that $b  $ {\em  mod} $  8 = 0$. Then, the
interpolation and scaling worst-case errors of
CVA, 1b, USA, 2s, 4s, 8s, 3LT and 4LT are the following:

\begin{center}
\begin{tabular}[h]{||c||c|c|c|c|c|c|c|c||}
\hline\hline
error/method & CVA & 1b & USA & 2s & 4s & 8s & 3LT & 4LT
\\ \hline
interpolation & $\frac{F \cdot b}{4}$ &  $\frac{F \cdot b}{4}$  &  $\frac{F
\cdot b}{4}$ &  $\frac{F \cdot b}{8}$ &
$\frac{F \cdot b}{16}$  &  $\frac{F \cdot b}{32}$ &  $\frac{F \cdot b}{16}$
&  $\frac{F \cdot b}{32}$
\\ \hline
scaling &   0 & 0& 0 & 0 &  $\frac{F \cdot b}{2^9}$  &  $\frac{F \cdot
b}{32}$  & $\frac{F \cdot b}{2^{12}}$ &   $\frac{F \cdot b}{2^7}$

\\ \hline
total & $\frac{F \cdot b}{4}$ &  $\frac{F \cdot b}{4}$  &  $\frac{F \cdot
b}{4}$ &  $\frac{F \cdot b}{8}$
& $\frac{F \cdot b}{16}$  &  $\frac{F \cdot b}{16}$ & $\frac{F \cdot b}{16}$
  &  $\frac{F \cdot b}{32}$
\\ \hline\hline
\end{tabular}
\end{center}
\end{theorem}

\begin{proof}
Let $b_M$ the size of the smallest sub-bucket produced by the method
$M$, where $M$ is either CVA, 1b, USA, 2s, 4s, 8s, 3LT or 4LT.
Observe that $b_M=b$ for CVA, 1b and USA (since they do not produce
sub-buckets), while $b_{2s}= \frac{b}{2}$, $b_M = \frac{b}{4}$ for
$M=$ 4s or $M=$ 3LT, $b_M = \frac{b}{8}$ otherwise.

Consider first the interpolation error
(by assuming that no scaling error occurs).

\noindent
{\bf Interpolation error bounds.}
It can be easily verified that the worst case for a method $M$ happens
whenever
both the following conditions hold:
\begin{enumerate}
\item [(1)]
there is a smallest sub-bucket, say $B$ (of size $b_M$) containing,
in the first half, $\frac{b_M}{2}$ frequencies with value $F$,
and, in the second half, $\frac{b_M}{2}$ frequencies with value 0, and
\item [(2)]
the range query involves exactly the first half of the sub-bucket $B$.
\end{enumerate}
The proof of this part is conducted separately for each method,
by determining the maximum absolute interpolation error:

\noindent{\bf CVA:}
In this case, $b_M = b$, that is the sub-bucket coincides with
the entire
bucket and the query boundaries are $1$ and $\frac{b}{2}$.
The cumulative value of the bucket is $F \cdot \frac{b}{2}$.
Under CVA, the estimated value of the query is $\frac{F \cdot
\frac{b}{2}}{b}\cdot \frac{b}{2}$,
that is $\frac{F \cdot b}{4}$. The actual value of the query is $\frac{F
\cdot b}{2}$.
Therefore the absolute error is $\frac{F \cdot b}{4}$.

\noindent{\bf 1b:}
We obtain the same absolute error $\frac{F \cdot b}{4}$.
Indeed, being the first value of the bucket $F$
(i.e., not null), 1-biased estimation does not give additional information
w.r.t. CVA.

\noindent{\bf USA:}
Also in this case, $b_M = b$, that is the sub-bucket coincides
with the entire
bucket and the query boundaries are $1$ and $\frac{b}{2}$.
The cumulative value of the bucket is $F \cdot \frac{b}{2}$.
USA assumes that the $\frac{b}{2}$ non null values are
located at equal distance from each other,
and each has the value $F$. As a consequence the estimated value of the
query
is $F \cdot \frac{b}{4}$, since the query involves just half non null
estimated values.
The actual value is $\frac{F \cdot b}{2}$. Thus, the absolute error is
$\frac{F \cdot b}{4}$, that is the same as CVA.

\noindent{\bf 2s:}
 In this case $b_M = \frac{b}{2}$.
According to the case CVA, the absolute error is
$\frac{F \cdot b_M}{4}$, that is
$\frac{F \cdot b}{8}$.

\noindent{\bf 4s} and {\bf 3LT}:
 Both 4s and 3LT produce sub-buckets
of size $\frac{b}{4}$. Thus, in these cases $b_M = \frac{b}{4}$.
Identically to the previous case, the absolute error is
$\frac{F \cdot b_M}{4}$, that is
$\frac{F \cdot b}{16}$.

\noindent{\bf 8s} and {\bf 4LT}:
Both 8s and 4LT produce sub-buckets
of size $\frac{b}{8}$. Thus, in these cases $b_M = \frac{b}{8}$.
Identically to the previous case, the absolute error is
$\frac{F \cdot b_M}{4}$, that is
$\frac{F \cdot b}{32}$.

Now we consider the scaling error.

\noindent{\bf Scaling error bounds.}
The proof that CVA, 1b, USA and 2s do not
produce scaling error is straightforward.
Let us consider the other methods:

\noindent{\bf 4s:}
Since each sub-bucket sum is encoded by 8 bits and is scaled
w.r.t. the overall bucket sum, the maximum scaling error is
$\frac{F \cdot b}{2^9}$.

\noindent{\bf 8s:}
Since each sub-bucket sum is encoded by 4 bits and scaled
w.r.t. the overall bucket sum, the maximum scaling error is
$\frac{F \cdot b}{2^5}= \frac{F \cdot b}{32}$.

\noindent{\bf 3LT:}
In this case, the scaling error may be propagated
going down along the path from the root to the leaves of the tree.
We may determine an upper bound of the worst-case error
by considering the sum of the maximum scaling error at each level.
Thus, we obtain the following upper bound:
$\frac{\frac{F \cdot b}{2^{12}} + \frac{F \cdot b}{2}}{2^{11}}$.
Indeed, the maximum scaling error of the first level is
$\frac{F \cdot b}{2^{12}}$. The above value is obtained by considering
that the maximum sum in the half bucket corresponding to the first level
is $\frac{F \cdot b}{2}$, and that going down to the second level
introduces a maximum scaling error obtained by dividing the
overall sum by $2^{11}$. Thus, the maximum scaling error for 3LT
is $\Theta(\frac{F \cdot b}{2^{12}})$ (that is, the scaling error of the first
level).

\noindent{\bf 4LT:}
For 4LT can be applied the same argumentation as 3LT, by
obtaining
that the maximum scaling error is of the same order as the first level.
That is, $\Theta(\frac{F \cdot b}{2^7})$, since the first level uses 6 bits.

The proof is thus completed.
\end{proof}

It is worth noting that,
as expected, 4LT and 8s produce the smallest interpolation worst-case error,
that is $\frac{F \cdot b}{32}$.
Considering also the results about scaling error,
the overall conclusion we may draw from the above analysis is that the best two
methods w.r.t. interpolation, that is 8s and 4LT, are not the same in terms of
scaling error.
Indeed 4LT shows a relevant accuracy improvement since the error
goes from $\frac{F \cdot b}{2^5}$ of 8s to $\frac{F \cdot b}{2^7}$
of 4LT.

In the next subsection we shall perform a number of experiments to
provide additional arguments in favor of the superiority of 4LT
estimation, by performing also an average-case analysis
of methods under a number of meaningful data distributions.
We shall not conduct experiments on the CVA
because we are aware that CVA uses 32 bits less and, therefore,
could reduce the size of the bucket, thus providing a better
accuracy. Actually, the performance analysis coincides with the one
of 2s estimation, that is CVA in half bucket.

\subsection{Experiments inside a Bucket}\label{sec-ExperimentsIntra}

In this section we report the results of a large number of
experiments performed with various synthetic data sets obtained
with different distributions. We measure the accuracy of all the
above mentioned methods in estimating range queries inside a
bucket. In particular, the methods considered are: USA, 1b, 2s, 8s, 3LT and
4LT. We observe that the space required for storing a bucket is the same
for all the considered methods.
Experiments are conducted
on synthetic data generated according several data distributions.
A data distribution is characterized by a distribution for frequencies and
a distribution for spreads.
Frequency set and value set are generated independently, then
frequencies are randomly
assigned to the elements of the value set.

\subsubsection{Test Bed.}

In this section we illustrate the test bed used in our
experiments. In particular, we describe (1) the
{\em data distributions}, that is the probability
distributions used for generating frequencies in the tested
buckets, (2) the {\em bucket populations}, that is the set
of parameters characterizing bucket used for
generating them under the
probability distributions, (3) the {\em data sets},
that is the set of samples produced by the combination of (1)
and (2), (4) the {\em query set and error metrics}, that is
the set of query submitted to sample data and
the metrics used for measuring the approximation error.

\noindent {\bf Data Distributions:} We consider four data
distributions: ({\bf 1}) {\em Zipf-$cusp\_max$ (0.5,1.0)}:
Frequencies are distributed according to a Zipf distribution
\protect\cite{Zipf49Human} with the $z$ parameter equal to $0.5$.
Spreads are distributed according to a Zipf {\em $cusp\_max$}
\protect\cite{Poo97} (i.e., increasing spreads following a Zipf
for the first half elements and decreasing spreads following a
Zipf distribution for the remaining elements) with $z$ parameter
equal to $1.0$. ({\bf 2})  {\em Zipf-$cusp\_max$(1.0,1.0).} ({\bf
3}) {\em Zipf-$cusp\_max$(1.5,1.0).} ({\bf 4}) {\em Gauss-rand}:
Frequencies are distributed according to a Gauss distribution with
standard deviation $1.0$. Spreads are randomly distributed as
well.

\noindent {\bf Bucket Populations:} A population is characterized
by the values of $c$ (overall cumulative frequency), $b$ (the
bucket size) and $t$ (number of non-null attribute values) and
consists of all buckets having such values.  We consider 9
different populations divided into two sets, that are called t-var
and b-var, respectively.

\noindent
{\em Set of populations t-var.}
It is a set of 6 populations of buckets, all of them with
$c=20000$ and  $b=500$. The 6 populations differ on the value of
the parameter $t$ ($t$=10, 100, 200, 300, 400, 500), and are denoted by
t-var(10), t-var(100), t-var(200), t-var(300), t-var(400) and
t-var(400), respectively.

\noindent
{\em Set of populations b-var.}
It is a set of 4 populations of buckets,  all of them with
$c=20000$. They  differ on the value of the parameters $b$ and
$t$. We consider $4$ different values for $b$ ($b$=100, 200, 500,
1000). The number of non-null values $t$ of each population is
fixed in a way that the ratio $t/b$ is constant and equal to
$0.2$; so the values of $t$ are 20, 40, 100 and 200. The four
populations are denoted by b-var(100), b-var(200), b-var(500) and
b-var(1000).

Moreover, a generic population whose parameter values are,
say, $\bar c$, $\bar b$ and $\bar t$ (for $c$, $b$ and $t$, respectively),
is denoted by p($\bar c$, $\bar b$, $\bar t$).

\noindent {\bf Data Sets:} As a data set we mean a sampling of the
set of buckets belonging to a given population following a given
data distribution. Each data set included in the experiments is
obtained by generating $100$ buckets belonging to one of the
populations specified above under one of the above described data
distributions. We denote a data set by the name of the data
distribution and the name of the population. For example, the data
set (Zipf-cusp\_max(0.5,1.0), b-var(200)) denotes a sampling of
the set of buckets belonging to the population of b-var
corresponding to the value 200 for the parameter $b$ following the
data distribution Zipf-cusp\_max(0.5,1.0).

We generate 23 different data sets
classified as follows:
(1)
{\bf Zipf-t} (i.e., Zipf data, different bucket density),
containing the five data sets (Zipf-cusp\_max(0.5,1), t-var($t$)), for
$t$=10,
100, 200, 300, 400, 500.
(2)
{\bf Zipf-b} (i.e., Zipf data, different bucket size),
containing the
four data sets (Zipf-cusp\_max(0.5,1), b-var($b$)), for $b$=100,
200, 500, 1000.
(3) {\bf Gauss-t} (i.e., Gauss data, different bucket density),
containing the five data sets (Gauss-rand, t-var($t$)), for $t$=10,
100, 200, 300, 400, 500.
(4)
{\bf Gauss-b} (i.e., Gauss data, different bucket size),
containing the
four data sets (Gauss-rand, b-var($b$)), for $b$=100, 200, 500,
1000.
(5)
{\bf Zipf-z} (i.e., Zipf data, different skew), containing the three
data sets Zipf-cusp\_max($z$,1.0), p(20000,400,200)), for $z$=0.5,
1.0, 1.5. Recall that p(20000,400,200) denotes the population characterized
by
$c=20000, b=400, t=200$.

Each class of data sets is designed for studying the dependence of
the accuracy of the various methods on a different parameter
(parameter $t$ measuring the density of the bucket, parameter $b$
measuring the size of the bucket and parameter $z$, measuring the
data skew). For each data set, 1000 different samples obtained by
permutation
of frequencies was generated and tested, in order to give
statistical significance to experiments.

\noindent {\bf Query set and error metrics:} We perform all the
queries $S[d]$, for all  $1 \leq d < b$. We measure the error of
approximation made by the various estimation techniques on the
above query set by using both:
\begin{itemize}
\item
the \em average \em of the \em relative
error \em $\frac{1}{b-1}\sum_{d=1}^{b-1}e_d^{rel}$,
where $e_d^{rel}$ is the \em relative error \em of the query with
range $d$, i.e., $e_d^{rel}=\frac{\vert{S[d]-
\widetilde{S}[d]}\vert}{S[d]}$, and

\item
the {\em normalized absolute error}, that is the ratio between the average
absolute error
and the overall sum of the frequencies in the bucket, i.e.
$\sum_{d=1}^{b-1}\frac{\vert{S[d]- \widetilde{S}[d]}\vert}{c \cdot b}$
\end{itemize}
where $\widetilde{S}[d]$ is the value of $S[d]$ estimated by the
technique at hand.

\subsubsection{Results of Experiments and Discussion.}

In this section we give a qualitative discussion about the approximation
error
of the considered methods, excluding USA and 1-biased, about which we have
already
provided a theoretical analysis in Section \ref{sec-Analysis}.
First we consider methods working simply by splitting the original
bucket, that are 2s, 4s and 8s.
For all these methods,
the estimation error may arise from the following approximation sources:

\begin{enumerate}

\item
the linear interpolation (i.e., CVA), concerning the evaluation of the query
inside
the ``smallest" sub-buckets (for instance, in the case of the 4s, the
smallest sub-buckets
are the quarts of the bucket),

\item

the numeric approximation, in case sums are stored by less than 32 bits
(note that only 2s is not affected by this error).

\end{enumerate}
We call error of type 1 and 2, respectively, the above described components
of the approximation error.

\subsubsection*{Relative error vs data density.}

Concerning error of type 1, what we expect is that, for all methods, it
increases as
data sparsity increases.
Indeed, in case of sparse data, the sum tends to concentrate in a few
points,
and this reduces the suitability of linear interpolation to approximate
the frequency distribution.
Moreover, we expect that such a component of the error
decreases as splitting degree increases:
for instance,
in case of 8s, which splits the bucket into 8 parts,
we expect more accuracy (in terms of the error of type 1) than
the 2s method. The reason is that having smaller sub-buckets
means applying linear interpolation to shorter
(and, thus, better linearly-approximable) segments of
the cumulative frequency distribution.

About error of type 2 we expect that both (i) it increases
as the splitting degree increases and (ii) it is independent of
data sparsity.
Claim (i) is explained by considering that increasing the splitting degree
means reducing the number of bits used for representing the sum of
sub-buckets.
Claim (ii) is related to the numeric nature of the error.

The observations above show the existence of a trade-off between the need of
increasing the splitting degree for improving CVA precision on one hand, and
the need of
using as more bits as possible for representing partial sums in the bucket
on the other hand.
However, we expect that such a trade-off is more evident in case of high
splitting degree,
that is, when the error of type 2 is more relevant.
For instance, recalling that the maximum absolute error of type 2 is
$\frac{c}{2^{k+1}}$,
where $k$ is the number of bits assigned to smallest sub-buckets, being
$k=4$ for 8s
and $k=8$ for 4s, the maximum absolute error of type 2 for 8s in
case $c=20000$
is 625 (i.e., about the 3\% of $c$) while it is 39 (i.e., a negligible
percentage of $c$) for 4s.

\begin{figure}[ht]
\begin{center}
\begin{tabular}{c}
\epsfig{file=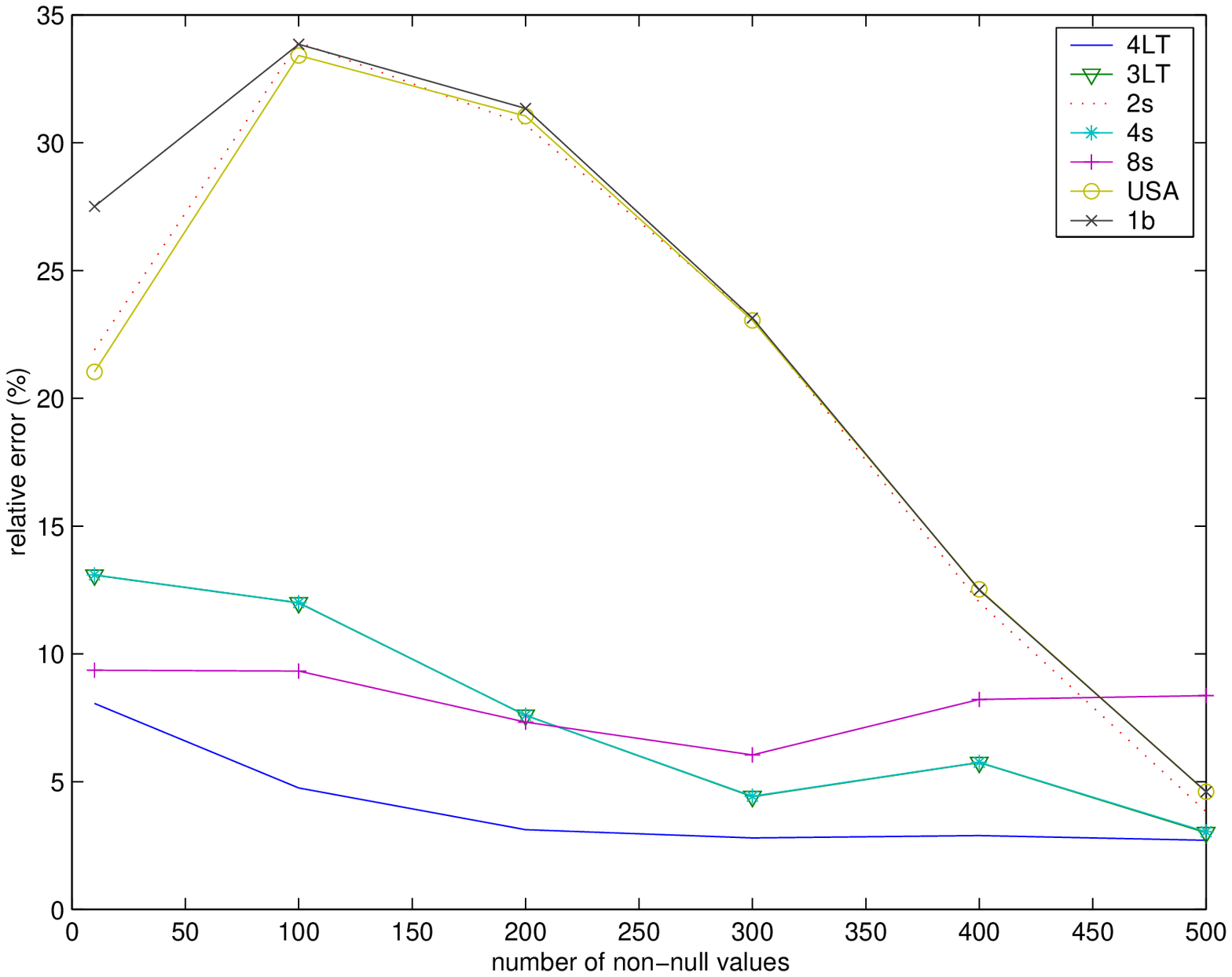,width=9cm} \\
{\bf (a)}: Error for different values of $t$ \\
\epsfig{file=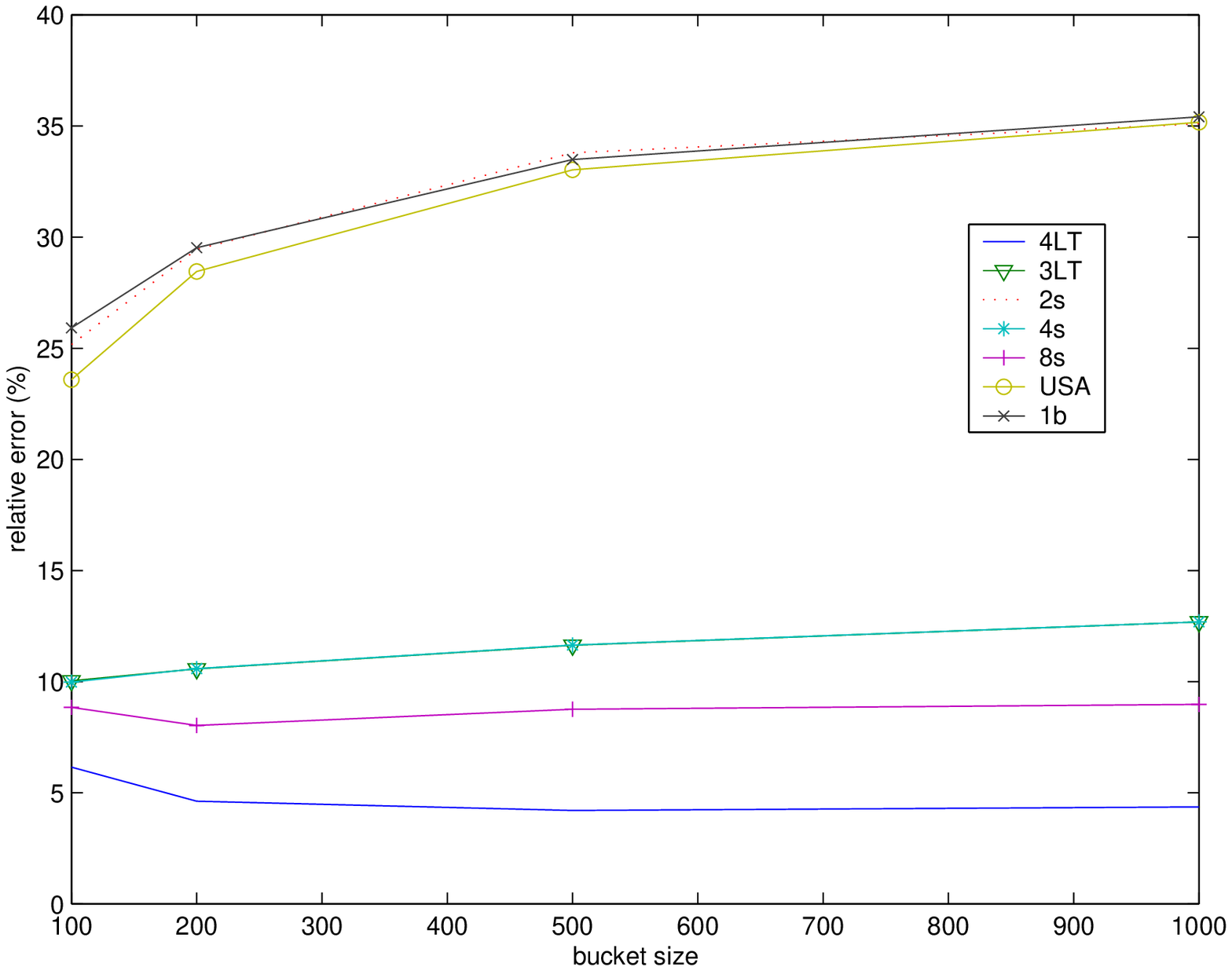,width=9cm} \\
{\bf (b)}: Error for different values of $b$
\end{tabular}
\end{center}
\caption{Experimental Results for data sets Zipf}\label{fig1}
\end{figure}

Experiments confirm the above considerations. By looking at graphs of
Figure \ref{fig1}.(a) we may observe that for 2s and 4s the error
decreases as the data density increases. On the contrary, for
8s, the error is quasi-constant (slightly increasing) in case of
Zipf distributions, while it is slightly decreasing (but much less
quickly than 4s) in case of Gauss distribution (see Figure
\ref{fig2}.(a)). Concerning the comparison between 2s, 4s and 8s,
we may observe in Figures \ref{fig1}.(a) that for low values of
data density, as expected, accuracy of 8s is higher than 4s and,
in turn, accuracy of 4s is higher than 2s. But, as observed above,
for increasing data density, trends of 4s and 8s suffer, in a
different measure, the presence of the error of type 2. This
appears quite evident in Figure \ref{fig1}.(a), whereby we may note
that 8s becomes worse than 4s from about 210 non null elements on
and the improving trend of 2s is considerable faster than the
other methods (since 2s does not suffer the error of type 2).

We observe that USA
gives better estimation than $1b$ on Zipf data (see Figures
\ref{fig1}.(a)). Accuracy of USA becomes the worst when the data sets
follow the Gauss distribution (see Figures \ref{fig2}.(a)).
This proves that the
assumption made by USA can be applicable for particular
distributions of frequencies and spreads, like those of data sets
Zipf-t. Results obtained on data sets distributed according a
Gauss distribution confirm the above claim: accuracy of USA
becomes the worst when the data sets have a random distribution as
it happens for Gauss-t (see Figure \ref{fig2}.(a)).

Concerning 1b we may observe that the behaviours of $1b$
and 2s are similar. As expected, the exploitation of the
information that the bucket is 1-biased does not give a
significant contribution to the accuracy of the estimation.
Indeed, the knowledge of the position of just one element in the
bucket does not add in general appreciable information.

Consider now the usage of the tree-indices 3LT and 4LT. Recall
that 3LT has the same splitting degree of 4s, since both methods
divide the bucket into 4 sub-buckets. Possible difference in terms
of accuracy between the two methods may arise from error of type 2.
Indeed, the tree-like organization of indices allows us to
represent the sum inside a given sub-bucket corresponding to a
node of the tree as a fraction of the sum contained in the parent
node, instead of the entire sum (as it happens for the "flat"
methods).
Thus, we expect that tree-indices produce smaller
errors of type 2. However, as previous noted, 4s produces a
negligible percentage of error of type 2. This explains why
3LT and 4s basically present
the same error (lines in the graphs are almost entirely
overlapped).

4LT has the same splitting degree as 8s (since both methods divide the
bucket into 8 sub-buckets). As a consequence, being appreciable the error of
type 2 of the 8s (as already discussed), we may expect
improvements by the usage of 4LT. This is that results from
experiments. 4LT has the best performances: it shows only benefits
deriving from the increasing of data density (producing the
reduction of error of type 1), with no appreciable increasing of
error of type 2. 4LT, thanks to the tree-like organization of the
sums, seems to solve the trade-off between increasing splitting
degree (for improving CVA precision) and controlling numeric error
arising from the usage of a reduced number of bits for
representing sums.

\subsubsection*{Relative error vs bucket size and
data skew.}

\begin{figure}[ht]
\begin{center}
\begin{tabular}{c}
\epsfig{file=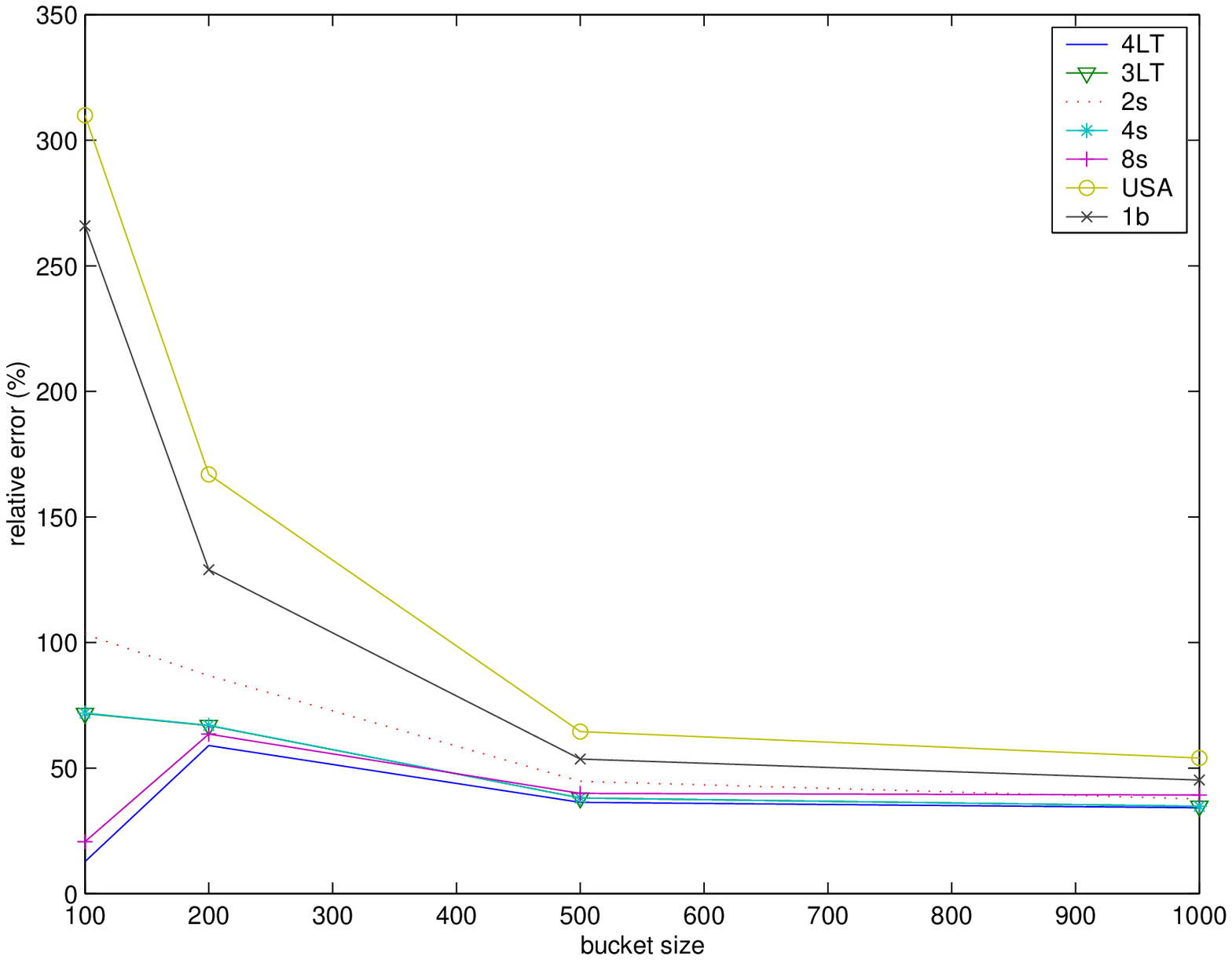,width=9cm} \\
{\bf (a)}: Data sets Gauss-t: error for different values of $t$ \\
\epsfig{file=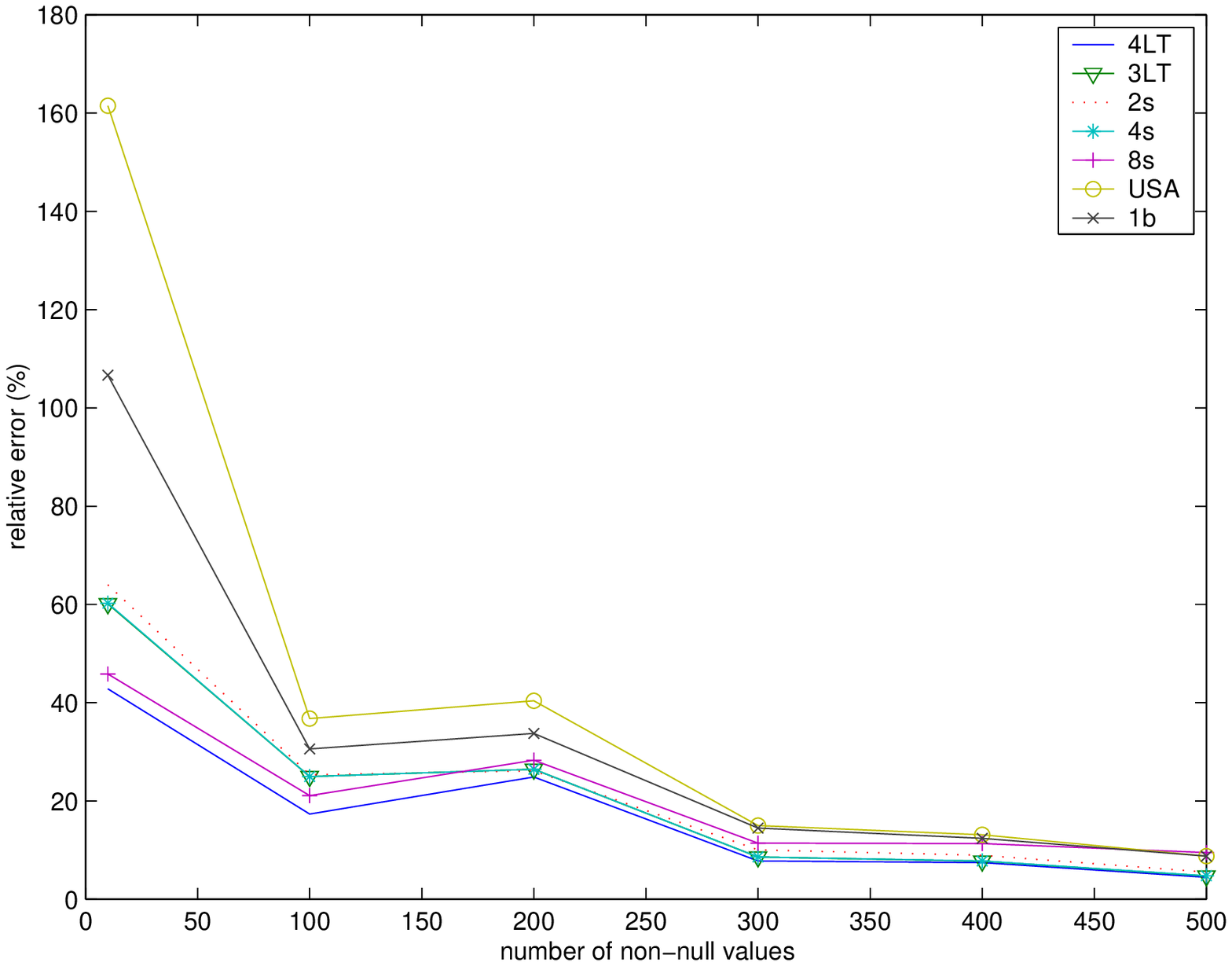,width=9cm} \\
{\bf (b)}: Data sets Gauss-D: error for different values of $b$
\end{tabular}
\end{center}
\caption{Experimental Results for data sets Gauss} \label{fig2}
\end{figure}

First consider populations b-var. Recall that for such data sets
we have maintained constant the data density around 20\%. Thus,
increasing the bucket size means increasing also non-null
elements. While, as for previous experiments, error of type 2 is
independent of the bucket size, (even though all the above considerations
about the relationship between error of type 2, splitting degree
and number of bits per smallest sub-buckets are still valid), we
expect that CVA precision suffers the variation of the bucket
size. Indeed, on the one hand the CVA precision decreases as the
bucket size increases, since, for a larger bucket, linear
interpolation is applied to a larger segment of the cumulative
frequency. But, on the other hand, increasing the bucket size means
increasing the number of non-null elements (keeping constant the overall sum)
and this means reducing the probability that the sum is
concentrated into a few picks. Thus, whenever the cumulative
frequency is smooth, linear interpolation tends to give better
results. Depending on data distribution, we may observe either
that the two opposite component compensate each other or one
prevails over the other. Indeed, experiments with Zipf data,
corresponding to Figure \ref{fig1}.(b), show that methods have a
quasi-constant trend (with a slight prevalence of the first
component), while experiments conducted on Gauss data,
corresponding to Figure \ref{fig2}.(b), show a net prevalence of the
second component (all the methods present a decreasing trend for
increasing bucket size). Such experiments do not give new
information about the comparison between the considered methods,
confirming substantially the previous results. Again 4LT has the
best performance.

\begin{figure}[h]
\begin{center}
\begin{tabular}{c}
\epsfig{file=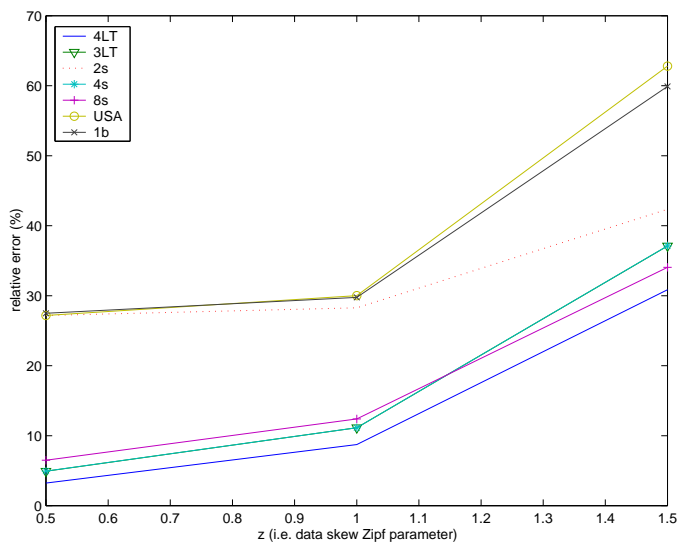,width=9cm}
\end{tabular}
\end{center}
\caption{Data sets Zipf-z: dependence on data skew} \label{fig3}
\end{figure}

Results of experiments conducted on the class of data sets Zipf-z,
for measuring the dependence of the accuracy of methods on the
data skew are reported in Figure \ref{fig3}. We note that all
methods become worse as $z$ increases (as it can be
intuitively expected).
The behaviours of $1b$ and 2s are similar, while 4LT shows the best
performance.

As a final remark we may summarize the comparison between the
considered methods concluding that the worst method is always 2s,
followed by 8s and then by 3LT and 4s for sparse data. On the
contrary, for dense data 3LT and 4s show better performance than
8s. Observe that 4s and 3LT have basically the same accuracy. The
best methods appears definitely 4LT.

\section{Applying the 4LT Index to the Entire Histogram}\label{sec-Improved}

The analysis described in the previous sections suggests to apply
the technique of the 4-level tree index to a whole histogram in
order to improve its accuracy on the approximation of the
underlying frequency set.
We stress that the problem
of investigating whether such an addition is really convenient
is not straightforward: observe that 4LT buckets use 32 bits more than CVA ones, and, then, for a fixed storage
space, allow a smaller number of buckets.
In this section we show how to combine
the 4LT technique with classical methods for constructing
histograms and we perform a large number of experiments to measure
the effective improvement given by the usage of the 4LT.
The advantage of the 4LT index is shown to be relevant also when
it is compared with buckets using CVA,
that is, when the storage space required by
4LT is larger than the original method.
Moreover, the 4LT index shows very good performances
if it is combines with a very
simple method for constructing histograms, called EquiSplit,
consisting on partitioning the attribute domain into equal-size
buckets.
Let us start with a quick overview of the most relevant methods
proposed so far for the construction of histograms.

\subsection{Methods for Constructing Histograms}

Besides the method used for approximating frequencies inside
buckets, the capability of a histogram of accurately approximating
the underlying frequency set strongly depends on the way such a
set is partitioned into buckets. Typically, criteria driving the
construction of a histogram is the  minimization of the error of
the reconstruction of the original (cumulative) frequency set from
the histogram. Partition rules proposed in
\protect\cite{Poosala96Improved,Jagadish98Optimal}, try to achieve
this goal. Among those, we sketch the description of two
well-known approaches: {\em MaxDiff } and {\em V-optimal} (see
\protect\cite{Poosala96Improved,Poo97} for an exhaustive
taxonomy). Note that these methods are defined for 2-histograms
but are in practice mainly used  for 1-histograms to minimize
storage consumption.

\noindent {\bf MaxDiff.} A MaxDiff histogram
\protect\cite{Cri81,Poosala96Improved} of size $h$ is obtained by
putting a boundary between two adjacent attribute values $v_i$ and
$v_{i+1}$ of $V$ if the difference between $f(v_{i+1}) \cdot
\sigma_{i+1}$ and $f(v_{i}) \cdot \sigma_{i}$ is one of the $h-1$
largest such differences (where $\sigma_i$ denotes the spread of
$v_i$). The product $f(v_{i}) \cdot \sigma_{i}$ is said the {\em
area} of $v_i$.

\noindent {\bf V-Optimal.} A V-Optimal histogram
\protect\cite{Poosala96Improved,Jagadish98Optimal} gives very good
performances. It is obtained by selecting the boundaries for each
bucket, $inf_i$ and $sup_i$, $1 \leq i \leq n$, so that
$\sum_{i=1}^n SSE_i$ is minimal, where $SSE_i =
\sum_{j=inf_i}^{sup_i} (f(j)-avg_i)^2$ and $avg_i$ is equal to the
average frequency in the $i$-th bucket, thus the cumulative
frequency in the whole bucket divided by the size $sup_i -
inf_i+1$.

We now propose to combine both methods, MaxDiff and V-Optimal,with
the 4LT index in order to  have an approximate representation of
frequency distributions inside the buckets. We shall compare the
so-revised methods with the original ones with CVA estimation at
parity of storage consumption. The results will show that the 4LT
index very much increases the estimation accuracy of both methods.
The additional estimation power carried by the 4LT index even
enables a very simple method like the one described below to
produce very accurate estimations.

\noindent {\bf EquiSplit.} The attribute domain is split into $k$
buckets of approximately the same size $b=\lceil m/k \rceil$. In
this way, as the boundaries of all buckets can be easily
determined from the value $b$, we only need to store a value for
each bucket: the sum of all frequencies. This method has been
first introduced in \protect\cite{Cri81} and, as the experimental
analysis will confirm, it has very good performances for low
skewed data, while its performances get worse in case of high
skew.

\subsection{Experiments on Histograms}\label{exp}

In this section we shall conduct several experiments both on
synthetic and real-life data in order to compare the effectiveness of
several histograms in estimating range query size.

\subsubsection*{Experiments on Synthetic Data.}
First we present the experiments performed on synthetic data.
Below we describe data sets, error metrics and the query set
considered in our experiments.

\noindent {\bf Available Storage:} Note that under CVA each bucket
stores only two integers, while with the 4LT index each bucket
needs  three integers. Assuming 32 bits the storage space for an
integer, given a fixed $K$ number of bits for the total storage
space required for the whole histogram, both MaxDiff and V-Optimal
under CVA produce $\lfloor \frac{K}{64} \rfloor$ buckets while
both of them with 4LT indices only produce $\lfloor \frac{K}{96}
\rfloor$ buckets. On the other hand, a bucket for EquiSplit just
needs one integer (the sum of all the frequencies), while for
EquiSplit-4LT it needs two integers. Thus, for a fixed $K$ number
of bits for the total storage space, EquiSplit with CVA produces
$\lfloor \frac{K}{32} \rfloor$ and EquiSplit with 4LT indices
produces $\lfloor \frac{K}{64} \rfloor$ as $MD\_CVA$.

For our experiments, we shall use a storage space, that is $42$
four-byte numbers to be in line with experiments reported in
\protect\cite{Poosala96Improved,Jagadish98Optimal}, which we
replicate. Using the above considerations, it can be easily
realized that MaxDiff with CVA, V-Optimal with CVA, and EquiSplit
with 4LT indices produce 21 buckets, EquiSplit with CVA produces
42 buckets, and both MaxDiff and V-Optimal with 4LT indices only
produce 14 buckets.

\noindent {\bf Data Distributions:} A data distribution is
characterized by a distribution for frequencies and a distribution
for spreads. Frequency set and value set are generated
independently, then frequencies are randomly assigned to the
elements of the value set. We consider 5 data distributions: ({\bf
1}) $D_1$: {\em Zipf-$cusp\_max$(0.5,1.0)}. ({\bf 2}) $D_2=$ {\em
Zipf-zrand(0.5,1.0)}: Frequencies are distributed according to a
Zipf distribution with the $z$ parameter equal to $0.5$. Spreads
follow a $ZRand$ distribution \protect\cite{Poo97} with $z$
parameter equal to $1.0$ (i.e., spreads following a Zipf
distributions with $z$ parameter equal to $1.0$ are randomly
assigned to attribute values). ({\bf 3}) $D_3=$ {\em Gauss-rand}:
Frequencies are distributed according to a Gauss distribution with
standard deviation $1.0$. Spreads are randomly distributed. ({\bf
4}) $D_4=$ {\em Zipf-$cusp\_max$(1.5,1.0)}. ({\bf 5}) $D_5=$ {\em
Zipf-$cusp\_max$(3.0,1.0)}.

\noindent {\bf Histograms Populations:} A population is
characterized by the value of three parameters, that are $T$, $D$
and $t$ and represents the set of histograms storing a relation of
cardinality $T$, attribute domain size $D$ and value set size $t$
(i.e., number of non-null attribute values).

\noindent
{\em Population $P_1$.}
This population is characterized by the following values for the
parameters: $D=4100$, $t=500$ and $T=100000$.

\noindent
{\em Population $P_2$.}
This population is characterized by the following values for the
parameters: $D=4100$, $t=500$ and $T=500000$.

\noindent
{\em Population $P_3$.}
This population is characterized by the following values for the
parameters: $D=4100$, $t=1000$ and $T=500000$.

\noindent
{\bf Data Sets:} Similarly to the experiments inside
buckets, each data set included in the experiments is obtained by
generating under one of the above described data distributions
$10$ histograms belonging to one of the populations specified
below. We consider the 15 data sets that are generated by
combining all data distributions and all populations.\\
All queries belonging to the query set below are evaluated over
the histograms of each data set:

\noindent
{\bf Query set and error metrics:} In our experiments, we use the
query set $\{X\leq d :d\in \U \}$ (recall that $X$ is the
histogram attribute and  $\U$ is its domain) for evaluating the
effectiveness of the various methods. We measure the error of
approximation made by histograms on the above query set by using
the \em average \em of the \em relative error \em
$\frac{1}{Q}\sum_{i=1}^Qe_i^{rel}$,
where $Q$ is the cardinality of the query set and $e_i^{rel}$ is
the \em  relative error \em, i.e.,
$e_i^{rel}=\frac{\vert{S_i-\widetilde{S}_i}\vert}{S_i}$,
where $S_i$ and $\widetilde{S}_i$ are the actual answer and the
estimated  answer of the query $i$-th of the query set.

\subsubsection{Results of the Experiments.} In Tables
\ref{table-1}, \ref{table-2} and \ref{table-3} the results of
experiments conducted on all data sets are reported. We denote the
methods MaxDiff, V-Optimal and EquiSplit with CVA by MD, VO and
ES, respectively; these methods with 4LT indices are denoted by
MD\_4LT, VO\_4LT, ES\_4LT.

\begin{table}
\begin{center}

\begin{tabular}[h]{|c|c|c|c|c|c|}
\hline\hline

$method/distr.$ &  $D_1$ &  $D_2$ &  $D_3$ & $D_4$ & $D_5$

\\ \hline

$ES$& $0.79$& $1.69$& $10.61$& $3.89$& $57.63$

\\ \hline

$ES\_4LT$& $0.29$& $0.84$& $2.01$& $2.89$& $29.63$

\\ \hline

$MD$& $4.29$& $19.37$& $11.65$& $7.02$& $31.46$

\\ \hline

$MD\_4LT$& $0.70$& $1.57$& $3.14$& $1.92$& $4.39$

\\ \hline

$VO$& $1.43$& $5.55$& $10.6$& $5.16$& $21.57$

\\ \hline

$VO\_4LT$& $0.29$& $1.33$& $2.32$& $1.62$& $3.15$

\\ \hline\hline

\end{tabular}

\end{center}

\caption{Pop. 1: error for various methods.}
\label{table-1}
\end{table}

\begin{table}
\begin{center}

\begin{tabular}[h]{|c|c|c|c|c|c|c|}
\hline\hline

$method/distr.$ &  $D_1$ &  $D_2$ &  $D_3$ & $D_4$ & $D_5$

\\ \hline

$ES$& $0.76$& $1.78$& $4.83$& $3.63$& $59.74$

\\ \hline

$ES\_4LT$& $0.28$& $0.84$& $6.40$& $1.40$& $31.12$

\\ \hline

$MD$& $5.79$& $16.04$& $6.65$& $13.56$& $33.51$

\\ \hline

$MD\_4LT$& $0.80$& $1.60$& $2.32$& $2.36$& $4.87$

\\ \hline

$VO$& $1.68$& $5.96$& $6.16$& $7.25$& $18.10$

\\ \hline

$VO\_4LT$& $0.32$& $1.41$& $4.85$& $1.53$& $3.12$

\\ \hline\hline

\end{tabular}

\end{center}

\caption{Pop. 2: error for various methods.}
\label{table-2}
\end{table}

\begin{table}
\begin{center}

\begin{tabular}[h]{|c|c|c|c|c|c|c|}
\hline\hline

$method/distr.$ &  $D_1$ &  $D_2$ &  $D_3$ & $D_4$ & $D_5$

\\ \hline

$ES$& $0.47$& $0.87$& $2.31$& $7.54$& $66.41$

\\ \hline

$ES\_4LT$& $0.27$& $0.35$& $1.14$& $3.59$& $25.01$

\\ \hline

$MD$& $8.37$& $2.89$& $3.30$& $3.46$& $25.01$

\\ \hline

$MD\_4LT$& $0.70$& $0.59$& $1.33$& $1.79$& $2.02$

\\ \hline

$VO$& $1.77$& $2.16$& $2.82$& $3.37$& $7.78$

\\ \hline

$VO\_4LT$& $0.32$& $0.56$& $1.24$& $1.68$& $1.82$

\\ \hline\hline

\end{tabular}

\end{center}

\caption{Pop. 3: error for various methods.}
\label{table-3}
\end{table}

\begin{figure}[h]
\begin{center}
\begin{tabular}{c@{\hspace{0.6cm}}c}
\epsfig{file=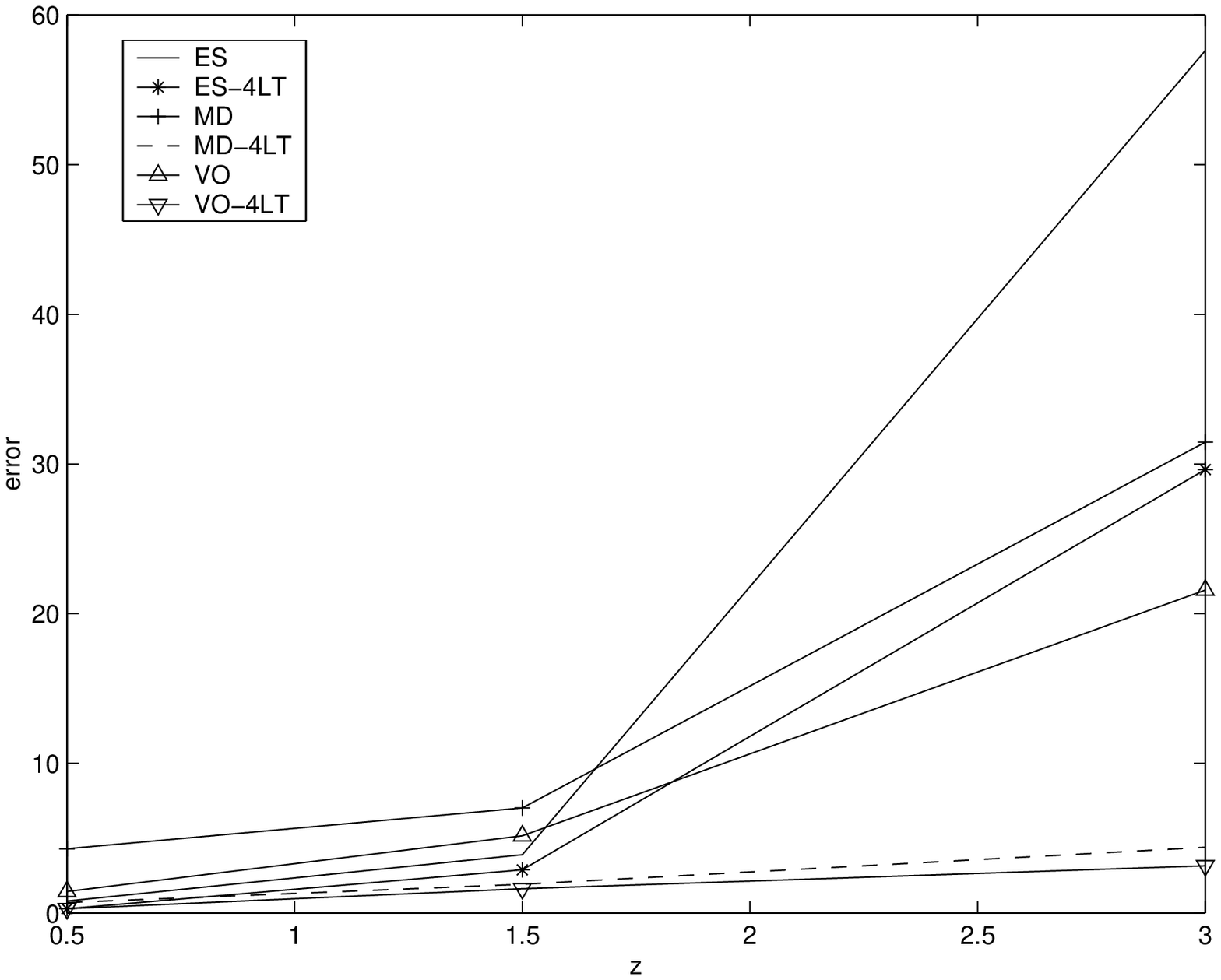,width=9cm} \\
{\bf (a)}: Dependence of the accuracy on the data skew \\
\epsfig{file=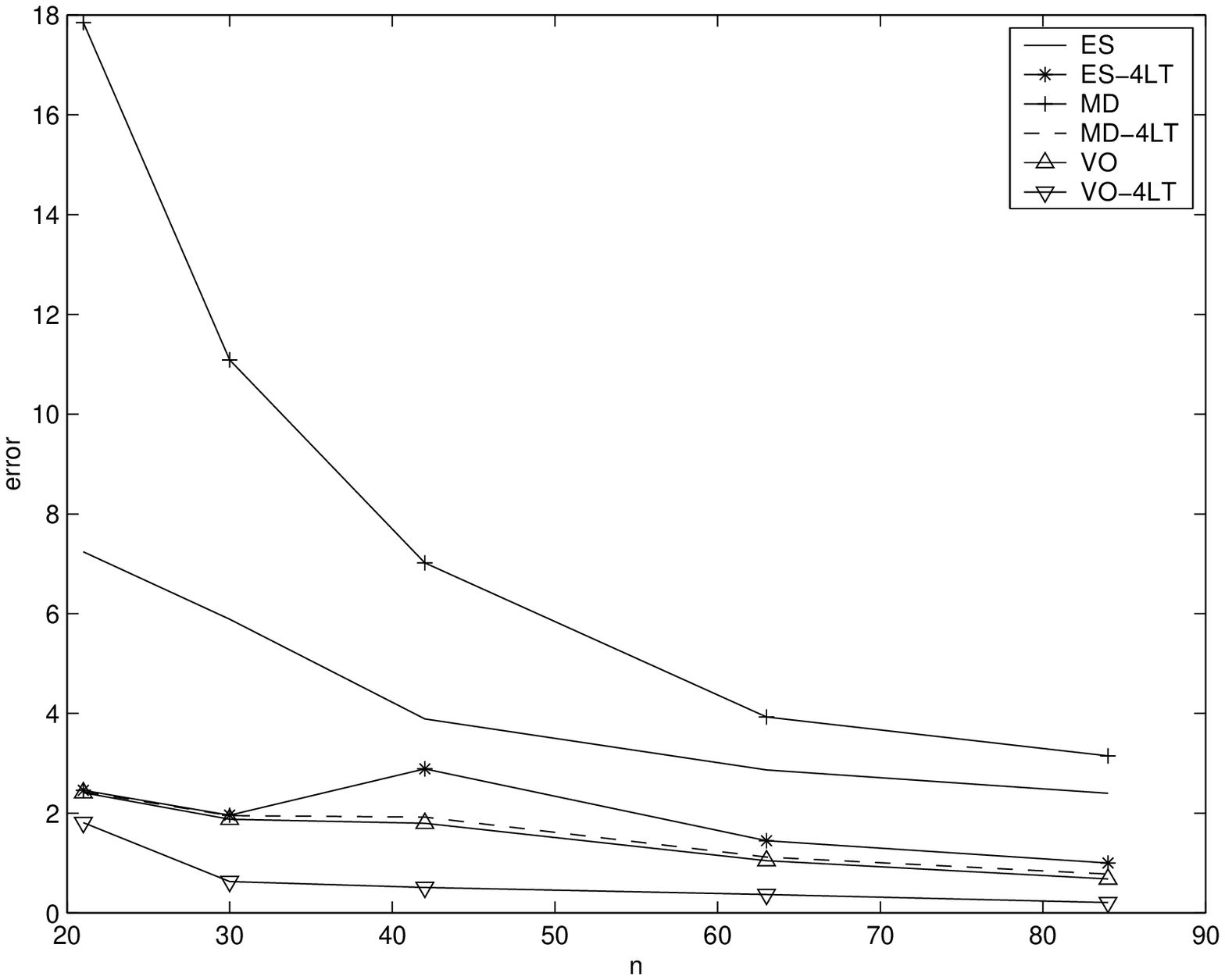,width=9cm} \\
{\bf (b)}: Dependence of the accuracy on the representation \\
  size (i.e., number of stored 4-byte integers)
\end{tabular}
\end{center}
\caption{Experimental Results}
\label{fig-5}
\end{figure}

The cross behavior of the various methods is
similar for the three populations. Experiments confirm the good
performance of the MaxDiff method and, particularly, of V-Optimal
but they also pinpoint that 4LT adds to both methods relevant
benefits. Indeed MD\_4LT and VO\_4LT show very low errors. Also
EquiSplit and EquiSplit-4LT have good performances. But, as shown
in Figure \ref{fig-5}.(a), where the dependence of the estimation
error on data skew is plotted, these methods quickly get worse for
high data skew. Indeed, in such cases, the benefit given by the
higher number of buckets is lost because of the high skew inside
buckets. In case of high skew, partition rules play a central
role, and the naive approach of EquiSplit is not suitable.
Interestingly, we observe that the improving of MaxDiff and
V-Optimal by the usage of 4LT indices is relevant also for high
skew, proving the effectiveness of such indices. In Figure \ref{fig-5}.(b)
we show the dependence of the accuracy of the methods on the amount of
space.
There, we consider the data distribution $D_4$ and the population
$P_1$ and generate 10 histograms belonging to $P_1$ according to
$D_4$ for different amounts of space. The aim of this experiment
is to study the behaviour of the various methods as the compression factor increases.
Clearly, when the available amount of space
increases, all methods behave well. The differences are more
relevant for values corresponding to high compression. Methods
using 4TL are the best. This can be intuitively explained by
considering that in case of large buckets the role of the
approximation technique inside buckets becomes more important than
the rules followed for constructing buckets.

\subsubsection*{Experiments on Real-Life Data.}
We have performed further experiments using real-life data. We
have considered two data sets (that we denote by Data Set A and
Data Set B) obtained from the {\em 1997 U.S. Census Statistics}
\protect\cite{Census}, by choosing two attributes of the table
{\em Special District Governments}, having the following
characteristics:

\noindent
{\bf Data Set A:}
attribute name: {\em Type Code},
domain size: $D= 998$,
number of non-null attribute values: $t = 787$,
cardinality: $T=34683$.

\noindent
{\bf Data Set B:}
attribute name: {\em Function Code},
domain size: $D= 99$,
number of non-null attribute values: $t = 32$,
cardinality: $T=34683$.

We use for each histogram the same amount of
storage space, that is $21$ four-byte numbers.
Query set and error metrics are the same used for experiments
on synthetic data.

\begin{table}

\begin{center}

\begin{tabular}{|c|c|c|}
  \hline
  method & data set A & data set B \\
  \hline
  $ES$ & 4.32 & 7.02 \\
   \hline
  $ES\_4LT$ & 0.97 & 3.59 \\
  \hline
  $MD$ & 11.30 & 22.82 \\
  \hline
  $MD\_4LT$ & 1.63 & 1.25 \\
  \hline
  $VO$ & 4.49 & 17.19 \\
  \hline
  $VO\_4LT$ & 1.86 & 3.05 \\
  \hline

\end{tabular}
\caption{Errors obtained on real data.}

\end{center}
\end{table}\label{realtable}

\noindent {\bf Results of the Experiments.} As shown in Table 4,
experiments on real data confirm the results obtained with
synthetic data. We note that 4LT adds to MaxDiff and V-Optimal
relevant benefits and both EquiSplit and EquiSplit-4LT have good
performances. Not surprisingly, for the data set A, EquiSplit-4LT produces the
smallest error. This can be explained
by considering that data of this set are rather uniform, and, in this case, as
discussed previously, the cheapest technique (in terms of storage space) gives the best
performances. In other words, the extra storage space required for recording
bucket boundaries of the more sophisticate techniques does not give benefits due to the
trivial data distribution.

\section{Conclusions}\label{sec-Conclusion}

In this paper we have presented a technique for improving the frequency estimation within
each bucket of a histogram. This technique goes beyond the simple methods used in the
literature, that is, the continuous value assumption and the uniform spread assumption.
Our method is based on the addition of a 32 data item to each bucket organized into a 4-level
tree index (4LT, for short) that stores, in a bit-saving approximate form, a number of hierarchical range queries
internal to the bucket. We have shown both theoretically and experimentally that such an additional
information effectively allows us to better estimate range queries inside buckets.
Interestingly, the usage of 4LT on top of histograms built through well-know techniques like
MaxDiff and V-Optimal, outperforms such histograms in terms of accuracy.
This claim is proven in the paper through a large number of experiments conducted on both synthetic
and real-life data, where classical histograms combined with 4LT are compared
with the standard versions (i.e., with no 4LT) under several
different data distributions at parity of consumed storage space.
It turns out that the price we have to pay
in terms of storage space by consuming 32 bits more per bucket
w.r.t. CVA-based histograms is overcome by the benefits given
by the improvement of precision in estimating
queries inside buckets.
Thus, the main conclusion we draw is that the 4LT index may represent a general technique
that can be combined with any bucket-based histogram for significantly
improving its accuracy.

{\footnotesize
\bibliography{isto}

\bibliographystyle{plain}
}

\end{document}